# FROG: Effective Friend Recommendation in Online Games via Modality-aware User Preferences

Qiwei Wang*
Tencent
Shenzhen, China
shimmerwang@tencent.com

Dandan Lin*
Shenzhen Institute of Computing Sciences
Shenzhen, China
lindandan@sics.ac.cn

Wenqing Lin†
Tencent
Shenzhen, China
edwlin@me.com

Ziming Wu
Tencent
Shenzhen, China
jimmyzmwu@tencent.com

## ABSTRACT

Due to the convenience of mobile devices, the online games have become an important part for user entertainments in reality, creating a demand for friend recommendation in online games. However, none of existing approaches can effectively incorporate the multi-modal user features (*e.g.*, images and texts) with the structural information in the friendship graph, due to the following limitations: (1) some of them ignore the high-order structural proximity between users, (2) some fail to learn the pairwise relevance between users at modality-specific level, and (3) some cannot capture both the local and global user preferences on different modalities. By addressing these issues, in this paper, we propose an end-to-end model FROG that better models the user preferences on potential friends. Comprehensive experiments on both offline evaluation and online deployment at Tencent have demonstrated the superiority of FROG over existing approaches.

**Source-code Availability:** Source code can be found at https://github.com/socialalgo/FROG.

## 1 INTRODUCTION

Due to the convenience of mobile devices, online games have become a significant component for user entertainments in reality [1, 16, 23–26, 28, 29, 35, 47–50]. In the online games, a player $u$ might want to connect with the other players for the purpose of sociality to interact with interesting users, or gaming requirements that encourage players to play the games together [23, 24, 47, 50]. However, it is difficult for player $u$ to search among billions or millions of players in the online game platforms, which has prompted a need for *friend recommendation* in the online games. Specifically, given a user (*i.e.*, player) $u$ in an online game platform, the friend recommendation task aims to recommend a *short* list of new potential friends with whom $u$ might be interested to have a connection. It also has been empirically verified that friend recommendation in the online games brings the growth of social network, and further increases the number of active interactions between users like game-playing or chatting [24, 50], leading to an increase of the total revenues in the game providers.

To tackle friend recommendation task, a straightforward approach is to utilize the natural graph structure behind the social networks where the nodes represent the users in a specific platform, and the edges between two nodes $u$ and $v$ denote that users $u$ and $v$ are friends in the platform. Based on the friendship graph, a plenty of *traditional proximity-based methods* can be adopted by firstly computing the user-user proximity scores based on the topological information, and then returning the top-$k$ users that have the highest proximity scores with respect to (w.r.t) a given user $u$ as the potential friends for user $u$, *e.g.*, Personalized PageRank [21, 23], the common friends-based triadic closure principle [6] or the path-based Katz centrality [18]. These proximity-based methods are based on the rationale that two users are more likely to connect if they have many common friends. However, this kind of methods *fail* to consider the valuable ***multi-modal*** information in the online games, *e.g.*, the user attributes and the in-game attributes, leading to inferiority in the performance. The multi-modal information has an important effect on users' decision to accept the recommendations. For example, a user whose profile image is a cartoon character is more likely to be interested in the users who have the cartoon profile image than those who have the profile image with natural scenery. Besides, advanced game players with high game level are more inclined to play with peers of comparable gaming ability, rather than those with vastly different game levels, as it leads to a more enjoyable gaming experience. Thus, it is equally important for friend recommendation to capture the user preferences by exploiting the multi-modal data.

Nevertheless, utilizing the multi-modal information for friend recommendation is not trivial due to the following two challenges.

- **How to jointly process the multi-modal data of different scales?** Multi-modality data can take various forms with different scales as they are collected from diverse domains. Although the use of user profile features has become pervasive in product/item recommendation applications [1] by learning the interactions between input features explicitly or implicitly, these *feature-based methods* [11, 13, 27, 33, 40, 51], *e.g.*, FM [33], DEEPFM [11] and AUTOINT [40], struggle to deliver comparable results when applied to friend recommendation tasks, since they not only *ignore the high-order structural proximity between users* revealed from the friendship graph, but also *fail to inject modality-aware signals*. While the state-of-the-art approach GRAFRANK [36] for friend recommendation exploits Graph Neural Network (GNN) to fuse the structural information, it *cannot effectively capture the pairwise relevance between users* since it focuses on the updating of individual user embedding.

---
*Co-first authors: Qiwei Wang, Dandan Lin.
†Corresponding author: Wenqing Lin.

| Model | EE | MP | HP |
|---|---|---|---|
| FM [33] | ✓ | ✗ | ✗ |
| DEEPFM [11] | ✓ | ✗ | ✗ |
| AUTOINT [40] | ✓ | ✗ | ✗ |
| GRAFRANK [36] | ✗ | ✗ | ✗ |
| FROG (**ours**) | ✓ | ✓ | ✓ |

**Table 1: Comparison of methods in terms of three abilities.**

- **How to effectively discriminate the effect of each modality for personalized recommendation?** Each modality contributes *differently* to a successful friend recommendation due to the following two reasons. (i) The information contained within each modality is distinct, leading to significant variations in their discriminative capabilities. (ii) The user preferences towards different modalities are *highly personalized*, *e.g.*, some users might focus on the profile images only, while some might be interested in the game-playing histories. Thus, it necessitates an effective methodology to discern the attentions of each user on each modality. Although previous work GRAFRANK [36] aggregates the neighbor information in a modality-specific manner, it *misses the global preferences* on different modality. The case becomes more challenging when combining the personalized attentions on modalities with the pairwise relevance between users.

To tackle these issues, we consider to develop an effective model for multi-modal friend recommendation in online games that have the following three abilities simultaneously. *(1) End-to-end learning (EE)*. The model learning should be end-to-end to generate the probability that any pair users will be friends for final friend recommendation, instead of emphasis on individual user embeddings. *(2) Modality-aware pairwise learning with high-order topological information (MP)*. The model can learn the implicit relationships between users at the fine-grained modality-specific level by incorporating the high-order structural proximity revealed in the friendship graph. *(3) Holistic personalized learning (HP)*. The model can learn the personalized user attentions on multi-modalities from both the *local* and *global* views. To the best of our knowledge, all of existing approaches in friend/item recommendation scenarios fail to satisfy above three key abilities *simultaneously*, as shown in Table 1. In this paper, we propose a novel end-to-end model for multi-modal friend recommendation in online games, termed FROG, that can fulfill the above three capabilities simultaneously. FROG has been deployed in the online games in Tencent and supports various friend recommendation scenarios. The following shows our **contributions**.

- We devise a Matching-Net inside FROG to learn the pairwise relationship between two users at the fine-grained modality-specific level. Specifically, for each pair of two users, it captures the implicit modality-aware signals by incorporating the high-order topological information revealed in the friendship graph.
- We combine a Local-Net and a Global-Net inside FROG for holistic personalized learning. In particular, our proposed Global-Net effectively discerns the attentions from the global perspective on all modalities by projecting the attention of each modality on a global sample decision plane.
- Comprehensive experiments demonstrated that the proposed model FROG significantly outperforms the state-of-the-art methods on two real datasets for friend recommendation, better than the second-best results **by up to 15.82%** and **14.59%** in terms of Hit-Rate and NDCG, respectively.
- We have developed FROG in a friend recommendation scenario of an online game in Tencent and conducted online A/B test to show its superiority over existing approaches.

## 2 PROBLEM DEFINITION

Given an online game with a massive number of users, let $V$ be the set of all users in the game. In the online game, each user $u \in V$ has a profile that contains the in-game attributes, such as game levels, personal descriptions, and profile images. Let $G(V, E)$ denote the friendship graph, where $V$ is the set of users and $E \subseteq V \times V$ is the set of friendship edges between users in $G$. In this paper, we only consider the *static* friendship graph that is captured before the daily model training phase. Let $n$ denote the number of users in the game platform before the daily model training. Besides, we used the following multi-modal user features in this paper.

**Multi-Modal User Features.** For each user $u \in V$, the following information from multi-modal is used: (1) *User-profile attributes*: It includes the in-game attribute data of a user $u$, *i.e.*, the game level, the gender, the active time in the game platform, and etc. (2) *Pairwise-interaction features*: It contains the interaction statistics between two users $u$ and $v$, *e.g.*, the number of common friends between $u$ and $v$, the number of times that $u$ and $v$ play game together, and etc. (3) *Profile images of users*: It are the profile images that each user $u$ displays in the game platform. (4) *In-game nicknames*: It is the textual nickname used by each user. (5) *Social networks*: It considers the existing in-game social properties of each user $u$, including the $K$-hop neighborhood of $u$ and the number of friends that $u$ has. In our experiments, $K = 2$.

Let $t$ be the number of modalities used for user features. The multi-modal data of $u$ is denoted by $X_u = [X_u^1, X_u^2, \ldots, X_u^i, \ldots, X_u^t]$, where $X_u^i$ is the data from the $i$-th modality of user $u$ and $i$ is an integer in $[1, t]$. The data from each modality might have different forms, *e.g.*, a vector or a graph (which will be elaborated in Section 3). We formally define the problem of *multi-modal friend recommendation* in a large-scale game platform as follows:

---
[1]The product/item recommendation task recommends a list of items to a given user.

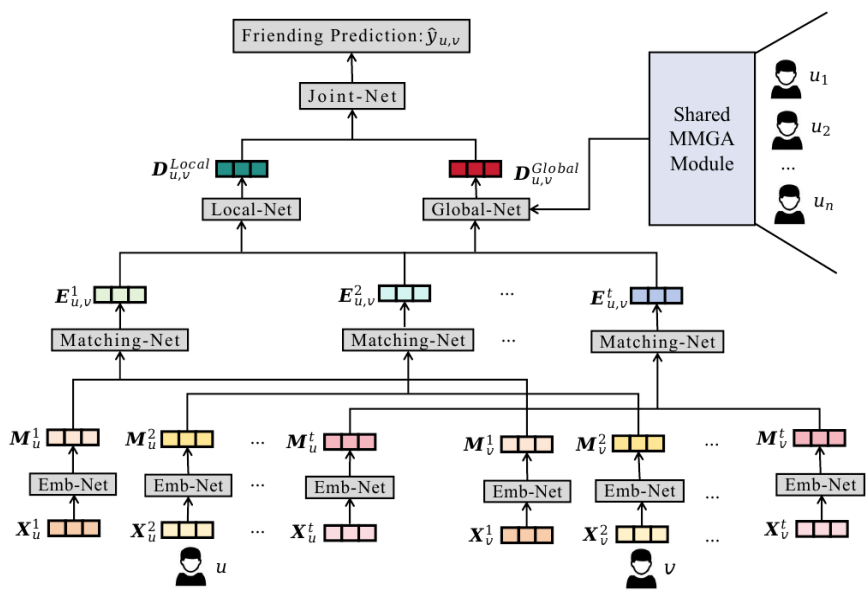

Figure 1: Framework of our proposed model FROG.

*Definition 1 (Multi-Modal Friend Recommendation).* Given the multi-modal user features $\mathcal{X}$ of all users in $V$, and a user $u \in V$, it recommends a list $L$ of $k$ users from $V$, such that the probability $\hat{y}_{u,v}$ of a user $v \in L$ that $u$ would establish a new friendship with $v$ is larger than other users that are not in the recommended list $L$, *i.e.*, $\hat{y}_{u,v} > \hat{y}_{u,w}$ for all $v \in L$ and $w \notin L$.

## 3 PROPOSED METHOD

### 3.1 Overview

Given two users $u$ and $v$, and their multi-modal data $\mathcal{X}_u$ and $\mathcal{X}_v$, our FROG generates the friending probability $\hat{y}_{u,v}$ that $u$ would establish a new friendship with $v$. Figure 1 shows the overall framework of our FROG that consists of five key components, *i.e.*, (1) the Emb-Net, (2) the Matching-Net, (3) the Local-Net, (4) the Global-Net, and (5) the Joint-Net. To be specific, FROG firstly feeds the multi-modal data $\mathcal{X}_u$ (w.r.t $\mathcal{X}_v$) into the Emb-Net to transform each modality of $\mathcal{X}_u$ (w.r.t $\mathcal{X}_v$) into a unified representation for the subsequent steps. Secondly, the obtained representations $\mathcal{M}_u$ (w.r.t $\mathcal{M}_v$) are coupled in the Matching-Net to learn the pairwise implicit similarity between $u$ and $v$ at the modality-specific level. Specifically, it jointly learns the mutual relevance between $u$ and $v$ for each modality, instead of regarding the modality data of each user individually. After that, the Local-Net and the Global-Net are processed simultaneously where the Local-Net facilitates the personalized preference of user $u$ on user $v$ and the Global-Net utilizes a shared multi-modal global attention mechanism (MMGA) to capture the global information of all users. Finally, the obtained local and global information is fused in the Joint-Net, which produces the friending prediction.

### 3.2 Details of FROG

In this section, we elaborate the five key components one by one.

**(1) Emb-Net.** The Emb-Net processes the multi-modality features into a unified representation. Specifically, given the multi-modal data $\mathcal{X}_u$ of a user $u$, for the $i$-th modality $X^i_u$, it generates a modality-specific embedding $M^i_u$ by considering the properties of the specific modality. After projecting each modality-specific embedding into a $d$-dimensional vector via three full-connected layers, it yields $\mathcal{M}_u$ as a new representation of $u$ by concatenating all vectors together, *i.e.*, $\mathcal{M}_u = \{M^1_u, \ldots, M^i_u, \ldots, M^t_u\}$ where $M^i_u$ is an $1 \times d$ vector.

In this paper, the multi-modal user features listed in Section 2 can be further categorized into the following three distinct types: (i) image data, *e.g.*, profile images of users; (ii) textual data, *e.g.*, in-game nicknames; and (iii) graph data, *e.g.*, social networks. In the followings, different techniques are tailored for each data type.

**Image data.** To obtain the embedding of each image, we train an image feature extractor IE in the offline phase. However, it is challenging to extract image features for online games due to two reasons: (a) the profile images are of large scale and (2) the volume of labeled image data is remarkably scarce. To solve these issues, our IE has two steps to improve the quality of extracting image features. At the first step, it adopts an existing model EFFICIENT-NET [41] on limited labeled image data that collected from both game platform and other public image sources. At the second step, it utilizes the latest unsupervised visual representation learning model MOCO [14] on large-scale unlabeled facial image dataset.

**Textual data.** To obtain the embeddings of each text, we directly employ the famous transformer-based encoder [19, 44].

**Graph data.** To obtain the high-order topological information in the friendship graph, we use the widely-used GNN model GRAPH-SAGE [12] to generate an embedding for each user by considering its $K$-hop neighborhood.

**(2) Matching-Net.** To satisfy the capability of modality-aware pairwise learning, the Matching-Net module jointly learns the pairwise similarity between two users on a specific modality. Specifically, given a pair of two users $u$ and $v$, and their representations $\mathcal{M}_u$ and $\mathcal{M}_v$ returned by the Emb-Net, the Matching-Net outputs a set $\mathcal{E}_{u,v}$ that contains the pairwise similarity embedding $\mathbf{E}^i_{u,v}$ between two users w.r.t the $i$-th modality, *i.e.*, $\mathcal{E}_{u,v} = \{\mathbf{E}^1_{u,v}, \ldots, \mathbf{E}^i_{u,v}, \ldots, \mathbf{E}^t_{u,v}\}$.

A straightforward approach to learn the pairwise similarity between two users is to compute the similarity between two representations $\mathcal{M}_u$ and $\mathcal{M}_v$ by using traditional $\mathcal{L}^p$-norm-based similarity measure, *e.g.* Euclidean Distance or Inner Product. However, this approach fails to discover the difference of mutual similarity. That is, the similarity score of user $u$ w.r.t $v$ is different from that of user $v$ w.r.t $u$ since $u$ and $v$ might have different preferences in making friends. To solve this problem, our Matching-Net module computes the pairwise similarity embedding by considering the mutual similarity. Formally, $\mathbf{E}^i_{u,v}$ is computed as follows:

$$\mathbf{E}^i_{u,v} = M^i_u \circ R^i_{u \to v} + M^i_v \circ R^i_{v \to u}, \quad (1)$$

where $R^i_{u \to v}$ and $R^i_{v \to u}$ are the $1 \times d$ relevance vectors of $v$ w.r.t $u$ and $u$ w.r.t $v$ on the $i$-th modality, respectively, and $\circ$ is Hadamard product that performs the element-wise multiplication of two vectors.

**Computation of relevance vectors.** To avoid being trapped in the local-level data, we utilize the attention mechanism to better model the mutual similarity. Specifically, the relevance vectors $R^i_{u \to v}$ and $R^i_{v \to u}$ are computed in three steps. Firstly, we measure the attention values $C^i_u$ (w.r.t $C^i_v$) of the $i$-th modality representation $M^i_u$ (w.r.t $M^i_v$) as below:

$$C^i_u = P^i_u M^i_u Q^i_u \text{ and } C^i_v = P^i_v M^i_v Q^i_v, \quad (2)$$

where $P^i_u$, $Q^i_u$, $P^i_v$ and $Q^i_v$ are learnable parameters, $P^i_u$ and $P^i_v$ are $d \times 1$ vectors, $Q^i_u$ and $Q^i_v$ are $d \times d$ matrices. Notice that $P^i_u$ and $Q^i_u$ work together to refine the granularity of the relationships inherent in $M^i_u$. Similarly for $P^i_v$ and $Q^i_v$. Secondly, the affinity matrix $G^i_{u,v}$

| Dataset | #Training | #Validation | #Testing | #Total |
|---------|-----------|-------------|----------|--------|
| Game1   | 697,979   | 174,495     | 3,210,833 | 4,083,307 |
| Game2   | 1,085,124 | 271,281     | 8,986,352 | 10,342,757 |

Table 2: Statistics of the datasets.

between $u$ and $v$ is computed by using the tanh function:

$$G_{u,v}^i = tanh\left((C_u^i)^T C_v^i\right). \quad (3)$$

Finally, the relevance vectors are calculated by using the mean-pooling on the affinity matrix:

$$R_{u \to v}^i = \sigma_1\left(\text{RowMean}\left(G_{u,v}^i\right)\right)^T, \quad (4)$$

$$R_{v \to u}^i = \sigma_1\left(\text{ColMean}\left(G_{u,v}^i\right)\right), \quad (5)$$

where $\sigma_1(\cdot)$ is an activation function.

**(3) Local-Net.** The Local-Net module explores the extent of how user $u$ is interested to be friends with user $v$ by considering the pairwise similarity from the perspective of each modality. Specifically, given the set $\mathcal{E}_{u,v}$ of pairwise similarity embeddings, a simple multi-layers perceptron (MLP) [13] is used to generate the local personalized preference as follows:

$$D_{u,v}^{local} = MLP\left(\phi_{1 \leq i \leq t}(\mathbf{E}_{u,v}^i)\right), \quad (6)$$

where $\phi$ is a element-wise concatenation function and the output of $\phi_{1 \leq i \leq t}(\mathbf{E}_{u,v}^i)$ is an $1 \times td$ vector. Since MLP can handle the interaction of features in the input data automatically, the locally discriminative information can be well observed.

**(4) Global-Net.** However, Local-Net learns the local personalized preference from a single training instance, which misses the global preference on different modalities. For example, the image data might be more important in the success of recommendation than the textual data, which can be revealed by all users. To address this issue, we propose the Global-Net module to inject the global user preferences on different modals by using a shared MMGA mechanism. To be specific, we use a trainable $1 \times d$ vector $A$ to represent a global training sample decision plane, namely, $A$ is treated as a global key shared with all training samples. Formally, given the set $\mathcal{E}_{u,v}$ of pairwise similarity embeddings, it calculates the global preference $D_{u,v}^{global}$ by projecting $\mathcal{E}_{u,v}$ on the global sample decision plane $A$ as follows:

$$D_{u,v}^{global} = \left(\sum_{1 \leq i \leq t} \mathbf{E}_{u,v}^i A^T\right) A. \quad (7)$$

With this paradigm, the global information provided by all users is integrated into the attention mechanism, augmenting the model with the capability of capturing the global pattern.

**(5) Joint-Net.** After obtaining both the local and global user preferences, the Joint-Net module is used to generate the friending probability $\hat{y}_{u,v}$ that $u$ would like to be friends with $v$, as follows:

$$\hat{y}_{u,v} = \sigma_3\left(W_2 \cdot \sigma_2\left(W_1 \cdot \phi(D_{u,v}^{local}, D_{u,v}^{global}) + b_2\right) + b_1\right), \quad (8)$$

where $\sigma_2$ and $\sigma_3$ are activation functions, $W_1$ and $W_2$ are the trainable weights while $b_1$ and $b_2$ are trainable biases.

### 3.3 Loss Function and Complexity Analysis

In online games, some users might accept or turn down several the friend recommendations, which should be well distinguished. To facilitate that, we exploit the FocalLoss [22] to construct the loss of prediction $\hat{y}_{u,v}$ as follows:

$$L = -\alpha (1-\hat{y})^\gamma log(\hat{y}) y - (1-\alpha) \hat{y}^\gamma log(1-\hat{y})(1-y), \quad (9)$$

where $y$ is the true label, $\hat{y}$ represents $\hat{y}_{u,v}$, $\alpha \in [0,1]$ and $\gamma \geq 0$ are hyper-parameters.

**Complexity.** Let $T_i$ be the unit cost for computing $M_u^i$ by Emb-Net for the $i$-th modality for user $u$. FROG takes $O(\sum_{i=0}^{t} T_i + td^3 + (td)^2 h + td + (h+d)^2)$ total time for each pair where $h$ is the number of dimensions of $D_{u,v}^{local}$.

Proof. For each pair $(u,v)$, It takes $O(\sum_{i=0}^{t} T_i)$ time to generate $\mathcal{M}_u$ and $\mathcal{M}_v$ by Emb-Net. Besides, it takes $O(td^3)$ time to generate $\mathcal{E}_{u,v}$ by Matching-Net since it takes $O(d^3)$ to compute $\mathbf{E}_{u,v}^i$ for the $i$-th modality by using the matrix multiplications and pooling computations. In addition, following [9], it takes $O((td)^2 h)$ to compute $D_{u,v}^{local}$ by Local-Net where $h$ is the number of dimensions of $D_{u,v}^{local}$, and takes $O(td)$ to compute $D_{u,v}^{global}$ by Global-Net since both $\mathbf{E}_{u,v}^i$ and $A$ are $1 \times d$ vectors. Finally, it takes $O((h+d)^2)$ to compute $\hat{y}_{u,v}$ by Joint-Net due to its linear computation. Thus, FROG takes $O(\sum_{i=0}^{t} T_i + td^3 + (td)^2 h + td + (h+d)^2)$ total time for each pair. □

## 4 EXPERIMENTS
### 4.1 Experiment Setup

**Datasets.** We used two *real* datasets collected from Tencent: (1) Game1 and (2) Game2. Table 2 shows the statistics of each dataset. *Collection processing.* For each dataset, it contains the friend-sending behaviors of the active users. Formally, in the game platform, a list of $k$ potential friends are recommended to a user $u$, where $k$ is a pre-defined parameter. If user $u$ sends a friending request to a recommended user $v$, then a pair $(u,v)$ is collected as *positive* instance into dataset; otherwise, a pair $(u,v)$ is collected as *negative* instance. We used the friend-sending behaviors and users' friending records interchangeably when it is clear in the context. To ensure privacy and confidentiality, the data used in experiments are strictly anonymous, following the ethical guidelines [31] by Tencent. Following previous work [11], for each dataset, we firstly collected the friending records of all active users in 7 consecutive days, and then randomly picked 80% friending history of each user as the training set and selected the remaining 20% as the validation set. Next, we collected all the friending records in the next 1 day as the testing set. Besides, due to the sparsity of the collected positive instances, we randomly selected 3 negative instances for each positive instance in the training and validation sets for better model training.

In total, Game1 has 4.08M of pairs while Game2 has 10.34M of pairs where M = $10^6$.

**Baselines.** We compared our method with *eleven* baselines including *Logistic Regression* (LR) [13], MLP [13] that consists of three fully-connected layers, *Factorization Machine* (FM) [33], DeepFM [11], AutoInt [40], AutoFIS+DeepFM [27] that removes redundant feature intersections inside DeepFM, DMF [15], SAGE+Max [12] that uses the element-wise *max pooling* in GraphSAGE model, SAGE+Mean [12] that uses the element-wise *sum-pooling* in GraphSAGE model, GraFRank [36] and EBR [39].

| Model | Game1 | | | | | | Game2 | | | | | |
|---|---|---|---|---|---|---|---|---|---|---|---|---|
| | HR@5 | NDCG@5 | HR@10 | NDCG@10 | HR@20 | NDCG@20 | HR@5 | NDCG@5 | HR@10 | NDCG@10 | HR@20 | NDCG@20 |
| LR | 0.2032 | 0.2534 | 0.2382 | 0.2663 | 0.2469 | 0.2777 | 0.0166 | 0.0124 | 0.0189 | 0.0131 | 0.0402 | 0.0210 |
| MLP | 0.2043 | 0.254 | 0.2395 | 0.2673 | 0.2483 | 0.2779 | 0.0167 | 0.0121 | 0.0191 | 0.0128 | 0.0406 | 0.0215 |
| FM | 0.209 | 0.2555 | 0.2404 | 0.2661 | 0.2477 | 0.2769 | 0.0145 | 0.0106 | 0.0163 | 0.0111 | 0.0357 | 0.0188 |
| DeepFM | 0.2067 | 0.2549 | 0.2401 | 0.2666 | 0.2487 | 0.2781 | 0.0149 | 0.012 | 0.0169 | 0.0126 | 0.0380 | 0.0208 |
| AutoInt | 0.2082 | 0.2558 | 0.2405 | 0.2669 | 0.2476 | 0.2782 | 0.0188 | 0.0148 | 0.0212 | 0.0155 | 0.0444 | 0.0237 |
| DMF | 0.2065 | 0.2543 | 0.2400 | 0.2654 | 0.2465 | 0.2772 | 0.0154 | 0.0115 | 0.0174 | 0.0121 | 0.0364 | 0.0198 |
| SAGE+Max | 0.2056 | 0.2573 | 0.2397 | 0.2656 | 0.2470 | 0.2772 | 0.0173 | 0.0136 | 0.0196 | 0.0143 | 0.0438 | 0.0225 |
| SAGE+Mean | 0.2074 | 0.2553 | 0.2404 | 0.2667 | 0.2475 | 0.2779 | 0.0176 | 0.0141 | 0.0199 | 0.0148 | 0.0439 | 0.0225 |
| GraFRank | 0.2088 | 0.2559 | 0.2400 | 0.2659 | 0.2475 | 0.2778 | 0.0182 | 0.0144 | 0.0204 | 0.0151 | 0.0438 | 0.0235 |
| EBR | 0.209 | 0.2562 | 0.2406 | 0.2668 | 0.2477 | 0.2781 | 0.0181 | 0.0142 | 0.0204 | 0.0149 | 0.0441 | 0.0235 |
| **FROG** | **0.2179** | **0.2637** | **0.2495*** | **0.2715*** | **0.2582*** | **0.2885*** | **0.0219** | **0.0163** | **0.0245*** | **0.0171*** | **0.0493*** | **0.0271*** |
| Improvement | 4.26% | 2.79% | 3.70% | 1.57% | 3.82% | 3.70% | 16.49% | 10.14% | 15.81% | 10.32% | 11.08% | 14.59% |

Table 3: Recommendation results of evaluated methods. The best and second-best results of each metric are highlighted in a bold font and underlined, respectively. The improvement is computed as the gains of the best result over the second-best result.

| Model | Game1 | | | | | |
|---|---|---|---|---|---|---|
| | HR@5 | NDCG@5 | HR@10 | NDCG@10 | HR@20 | NDCG@20 |
| without Matching-Net | 0.2129 (2.3%↓) | 0.2587 (1.9%↓) | 0.2415 (3.31% ↓) | 0.2669 (1.72% ↓) | 0.2487 (3.82% ↓) | 0.2787 (3.52% ↓) |
| without Local-Net | 0.2113 (3.1%↓) | 0.2553 (3.3%↓) | 0.2395 (4.18% ↓) | 0.2657 (2.18% ↓) | 0.2471 (4.49% ↓) | 0.2774 (4.00% ↓) |
| without Global-Net | 0.2127 (2.4%↓) | 0.2572 (2.5%↓) | 0.2407 (3.66% ↓) | 0.2673 (1.57% ↓) | 0.2489 (3.74% ↓) | 0.2777 (3.89% ↓) |
| FROG | **0.2179** | **0.2637** | **0.2495** | **0.2715** | **0.2582** | **0.2885** |

Table 4: The effect of each module in FROG for friend recommendation performance on Game1 dataset.

| Model | accRate | intRate |
|---|---|---|
| AutoInt | 42.86% | 50.01% |
| **FROG (Ours)** | **57.14%** | **62.51%** |

Table 5: Results of A/B testing in an online game of Tencent.

**Default parameters.** We exploit the Adam optimizer for training the models. Besides, we set the learning rate as 0.001, the max epochs as 50, and the batch size as 1024. For activation functions used in FROG, $\sigma_1$, $\sigma_2$ and $\sigma_3$ are the softmax ReLu and Sigmoid functions, respectively. For a fair comparison, we used the same embeddings of multi-modal data obtained by Emb-Net and the same training strategy for both our model and the competitors. Besides, for each evaluated method, we choose the model that performs the best in the validation set, to be evaluated for the testing set.

**Environment.** We run the experiments on a machine with a Tesla V100 GPU, 22 CPU cores, and 90 GB shared CPU memory. We implemented each evaluated method by using TensorFlow.

**Evaluation metrics.** Following previous work [36], we used two widely-used metrics to evaluate the performance of the proposed approach, i.e., *Hit-Rate* (HR@$k$) and *Normalized Discounted Cumulative Gain* (NDCG@$k$), where $k$ is varied from {5, 10, 20}. For each experiment, we repeated 5 times and reported the average results.

## 4.2 Experimental Results

**Overall performance.** Table 3 shows the results on two datasets where the best and second-best results of each metric are highlighted in a bold font and underlined, respectively. From the results, we can see that our proposed model FROG achieves the best performance on all datasets when $k$ is varied from 5 to 20. To be specific, on the largest dataset Game2, FROG has better recommendation performance than the second-best baseline AutoInt **by up to 15.82%** and **14.59%** in terms of HR and NDCG, respectively. It is because FROG considers both the pairwise modality-aware signals between users while AutoInt fails. Moreover, compared with GraFRank that utilizes both the multi-modalities and the social topology information, the performance of FROG is higher no matter how $k$ is changed, showing the effectiveness of FROG by considering both the local and global personalized user preferences on different modalities.

**Ablation study.** We evaluated the effect of each module in FROG with its three degraded variants. Table 4 shows the results on the Game1 dataset by varying $k$ from 5 to 20. It demonstrates that each module has the essential influence on the friend recommendation, *namely*, FROG using all three modules has the best performance **by up to 4.49%** and **4.00%** in terms of HR and NDCG, respectively.

**Online deployment.** We deployed it in an online *First Personal Shooter* (FPS) game of Tencent, which is is a multiplayer online game with billions of users. We selected AutoInt as the competitor since it achieves the second-best results in most of cases.

We deployed our proposed model FROG and AutoInt in an in-house cluster with hundreds of machines, each of which has 16GB memory and 12 Intel Xeon Processor E5-2670 CPU cores. To make the models updated with the frequently changing data in the game, we re-build the models from scratch every 12 hours. Besides, each model generates 10 potential friends for each list of recommendations for each user in the platform. For each experiment, we run the algorithms for a consecutive 15 days where nearly 29 millions of users are involved.

In the experiments, we have two groups, i.e., control group applying AutoInt model and the treatment group using the proposed model FROG. We conduct the online A/B test by assigning each user to a randomly chosen group.

*Metrics.* We used two evaluation metrics *i.e.*, *the acceptance rate* (accRate) that user accepts the recommendations and *interaction rate* (intRate) that user interacts with the newly-connected friends after acceptance. Specifically, the accRate is formally defined as below:

$$\text{accRate} = \frac{\#\text{successRec}}{\#\text{totalRec}},$$

where #successRec and #totalRec denote the number of successful new friendships made by the friend recommendation service in the game and the total number of friend recommendations exposed to users, respectively. And the intRate is formally defined as below:

$$\text{intRate} = \frac{\#\text{interactRec}}{\#\text{successRec}},$$

where #interactRec denotes the number of the newly-constructed friendships that have further interactions in the game.

These two metrics ensures both the accurate predictions and the quality of the predictions of each model.

*Results.* Table 5 shows the results of A/B tests. FROG outperforms AutoInt by up to **33.3%** and **25.5%** in terms of both accRate and intRate, respectively, showing the effectiveness of FROG in reality.

## 5 RELATED WORK

In this section, we briefly review existing methods about friend recommendation tasks, general user-item recommendation tasks and multi-modal learning techniques.

**Friend recommendation.** Existing friend recommendation methods can be classified into three categories: (1) *traditional proximity-based* methods [6, 20]; (2) *ranking-based* methods, which consider the friend recommendation problem as a learning-to-rank task [5]; and (3) *embedding-based* methods [10, 30, 36, 39], which compute the user embeddings by utilizing either the *graph embedding-based* approaches or the popular *graph neural network-based* approaches.

*(1) The traditional proximity-based methods.* These methods have been introduced in Section 1. For the space limit, we omit here.

*(2) The ranking-based methods.* For the ranking-based methods [5, 32, 34], they rank the friendship probability of observed friends of each user to be larger than that of the unobserved ones. Among these methods, BayDNN [5] is the latest one that defines the pairwise ranking relationships between two users by combining the *Bayesian Personlized Ranking* (BPR) model [34] and Deep Neural Networks (DNN). However, they do not consider the high-order topological information in the friendship graph.

*(3) The embedding-based methods.* Given a graph, the graph embedding based approaches learn latent node representations to capture the structural properties of a node and its neighborhoods, *e.g.*, Node2Vec [10] and DeepWalk [30], which are easily to extended for friend recommendations in the social networks. However, these approaches suffer from the expensive time cost for training on a large-scale graph [36]. To tackle this issue, the GNN-based approaches [36, 39] were widely-used for real-world scenarios by efficiently learning the user embeddings using GNN models, *e.g.*,

GraFRank [36] which generates user embeddings by using both the multi-modal user features and the topological information in graph, and EBR [39] which enriches the friend candidates beyond traditional 2-hop neighborhoods by using the graph-aware user embeddings with GraphSAGE [12]. However, EBR fails to enrich the multi-modal user features, while GraFRank misses the personalized impact of different modalities on a specific user.

Although several models [7, 37, 38, 42] utilize the social network as an auxiliary data source to model user behavior in social platforms and improve quality of item recommendations to users, they cannot be adapted well to the friend recommendation task since they fail to facilitate creating a better social network of users [36].

**General user-item recommendation.** It is straightforward to adapt the approaches on general user-item recommendation for the friend recommendation problems since both problems can be regarded as *Click-through Rate* (CTR) prediction [52] that aims to estimate the probability that a user will click on a certain item (*i.e.*, a user in this paper). Hence, we briefly review a list of existing work for user-item recommendation in below. FM [33] projects each feature into a low-dimensional vector and models feature interactions by inner product, which works well for sparse data. Due to the effectiveness of deep learning methods, several models use MLP to improve FM, *e.g.*, Wide &Deep [4] and DeepFM [11], which simply enumerate all the second-order feature interactions where most of interactions are useless and noisy. To prune the unnecessary interactions between features, several works were proposed, *e.g.*, AutoInt [40] that explicitly models feature interactions with attention mechanism, and AutoFIS [27] that automatically identifies important feature interactions. Besides, instead of removing the useless feature interactions, a recent work FINAL [51] explores a better alternative to the MLP backbone that could potentially replace MLPs. However, all of these work fails to tackle the friend recommendation well since all of them not only ignore the high-order structural information but also fail to inject modality-aware signals.

**Multi-modal learning-based recommendation.** Due to the fact that the information from different modalities has different influence on users for decision making, the fusion of multi-modal features, *e.g.*, texts, images, and graphs, has been widely-used in plenty of models [8, 15, 17, 36], *e.g.*, DMF [15] and GraFRank [36]. However, these methods either consider the modalities as *independent data sources* or fails to learn both *local and global effect of modalities* for friend recommendation. In this work, we model the user preferences from both local and global perspective. As shown our experiments in Section 4, our design of local (*resp.* global) module can increase the performance of friend recommendation **by up tp 23.74%** (*resp.* **18.34%**) in terms of NDCG @20.

Besides, existing methods [2, 3, 43, 45, 46] that utilize multi-modal information for item recommendation, also cannot be used in our studied problem since they either (1) fail to consider modality-aware information [46], or (2) fail to consider the pairwise user preferences [2, 43, 45] , and or (3) focus on the alignment of multi-modal information from a source domain to a target domain [3].

## 6 CONCLUSION

In this paper, we propose an end-to-end model FROG for multi-modal friend recommendation in online games. The model focuses on the modality-aware pairwise learning from both local and global user preferences by utilizing the multi-modal user features. Comprehensive experiments have demonstrated its effectiveness for friend recommendation scenarios in real-world online games.